\documentclass[twocolumn,showpacs,preprintnumbers,amsmath,amssymb, pr]{revtex4}
\usepackage{mathrsfs}

\usepackage{graphicx}
\usepackage{dcolumn}
\usepackage{bm}
\usepackage{epsfig}

\begin{document}

\title{Phase Dynamics in Intrinsic Josephson Junctions and its Electrodynamics}

\author{Shizeng Lin\(^{1,2}\) and Xiao Hu\(^{1,2,3}\)}
\affiliation{\(^{1}\)WPI Center for Materials
Nanoarchitectonics, National Institute for Materials Science, Tsukuba 305-0047, Japan\\
\(^{2}\)Graduate School of Pure and Applied Sciences, University of
Tsukuba, Tsukuba 305-8571, Japan\\
\(^{3}\)Japan Science and Technology Agency, 4-1-8 Honcho,
Kawaguchi, Saitama 332-0012, Japan}

\date{\today}

\begin{abstract}
We present a theoretical description of the phase dynamics and its corresponding electrodynamics in a stack of inductively coupled intrinsic Josephson junctions of layered
high-$T_c$ superconductors in the absence of an external magnetic field. Depending on the spatial structure of the gauge invariant phase difference, the dynamic state is
classified into: state with kink, state without kink, and state with solitons. It is revealed that in the state with phase kink, the plasma is coupled to the cavity and the plasma
oscillation is enhanced. In contrast, in the state without kink, the plasma oscillation is weak. It points a way to enhance the radiation of electromagnetic from high-$T_c$
superconductors. We also perform numerical simulations to check the theory and a good agreement is achieved. The radiation pattern of the state with and without kink is
calculated, which may serve as a fingerprint of the dynamic state realized by the system. At last, the power radiation of the state with solitons is calculated by simulations. The
possible state realized in the recent experiments is discussed in the viewpoint of the theoretical description. The state with kink is important for applications including
terahertz generators and amplifiers.
\end{abstract}


\maketitle
\section{Introduction} The electromagnetic waves in the terahertz(THz) region,
which is defined in the range from $0.1$THz to $10$THz, have wide applications, such as drug detection, materials
characterization, security check, and so on. This has sparked considerable efforts to seek compact and low-cost
solid-state generators.\cite{Ferguson02,Tonouchi07}

It has been known for a long time that Josephson junctions can be used as electromagnetic oscillators. The power
radiated from a single junction, however is in the range of $\rm{pW}$, which is too small for practical applications.
The frequency is about one hundred gigahertz because of the small superconducting energy gap for conventional
superconductors.\cite{Yanson65,Dayem66,Zimmerma66,Pedersen76} Although one may integrate a large array of Josephson
junctions made of conventional superconductors on a chip to enhance the radiation
power,\cite{Finnegan72,Jain84,Durala99,Barbara99} the frequency is still below terahertz. The discovery of intrinsic
Josephson junctions in layered high-$T_{c}$ superconductors, such as $\rm{Bi_2Sr_2CaCu_2O_{8+\delta}}$(BSCCO) provides
a very nice candidate for terahertz oscillator.\cite{Kleiner92} The advantages of intrinsic Josephson junctions over
conventional low temperature junctions are as follows. First, the junctions are homogeneous in the atomic scale, which
makes the coherent radiation in large number of junctions possible. Secondly, the energy gap is about $60$meV which
corresponds to $15$THz. The terahertz Josephson plasma if excited is thus free from Landau damping.\cite{Tachiki94}

One idea to excite the terahertz wave inside the intrinsic Josephson junctions is by the motion of Josephson vortices lattice induced by an in-plane magnetic field and a transport
current, which has already been investigated both theoretically and experimentally.\cite{Koyama95, Tachiki05,Wang06, Bulaevskii06, Kadowaki06,Bae07,szlin08} In spite of these
works, it is still lack of clear evidence of coherent radiation. Radiation from BSCCO with injection of quasiparticles has been
reported\cite{Shafranjuk99,Kume99,Lee00,Iguchi00a,Iguchi00b}. Alternatively, the terahertz radiation without a magnetic field has also been attempted\cite{Batov06,Bulaevskii07,
Ozyuzer07,Koshelev08, kadowaki08}. Recently, a strong coherent radiation from BSCCO in the absence of a magnetic field was observed\cite{Ozyuzer07, kadowaki08}, where the mesa of
the single crystal of BSCCO forms a cavity. The breakthrough in the experiments has inspired considerable theoretical and experimental efforts, aiming to reveal the mechanism of
strong radiation.\cite{szlin08b,Koshelev08b,Mastumoto08,Tachiki08,Hu08,Wang09} A new dynamic state has been suggested to explain the experiments\cite{szlin08b,Koshelev08b}.

It is well known that there exist various dynamic states, such as the McCumber state and states with solitons, with different \emph{IV} characteristics in a Josephson
junction\cite{Kleiner00,szlin08b} due to the nonlinearity. In the McCumber state, the gauge invariant phase difference is uniform in space, which we will refer to as the state
without kink in later discussions. The soliton solutions are also well known, especially in a single junction, where a quantized particle-like object of $2\pi$ phase variation
travels along the junction. Recently, a new dynamic state was found, where a $(2m+1)\pi$ phase kink is localized inside the junction with an integer
$m$.\cite{szlin08b,Koshelev08b} We will refer to this state hereafter as state with kink. Because of the complexity in the dynamics in this highly nonlinear system, theoretical
understanding of phase dynamics and its electrodynamics, and finding the optimal state are expected to be helpful for experimental realizations of terahertz generators.
Assessments of such a device from a theoretical aspect are also needed.

For this purpose, in this paper we provide more details on the new dynamic state found in the previous
study.\cite{szlin08b} In Section II, we first derive the Lagrangian of a stack of Josephson junctions based on the
superconductor-insulator-superconductor(SIS) model. From the Lagrangian we derive the inductively coupled sine-Gordon
equation. It also gives the power balance condition. In section III, we develop a general procedure to solve the
coupled sine-Gordon equation from the spectrum analysis, from which we derive the solutions with and without phase
kink. We calculate analytically the \emph{IV} characteristics and power radiation of the state with and without kink
when the plasma oscillation is small. Numerical simulations are performed to verify the analytical solutions and a good
agreement is attained. In section IV, the energy stored in the system is evaluated. In section V, the far-field
radiation pattern from the mesa is calculated both at state with and without kink, which can be used to distinguish the
different dynamic states. In section VI, the power radiation from the state with solitons is investigated by
simulations for comparison. At last, the paper is concluded with a short discussion.

\section{Lagrangian and Model Equation}
\begin{figure}[t]
\psfig{figure=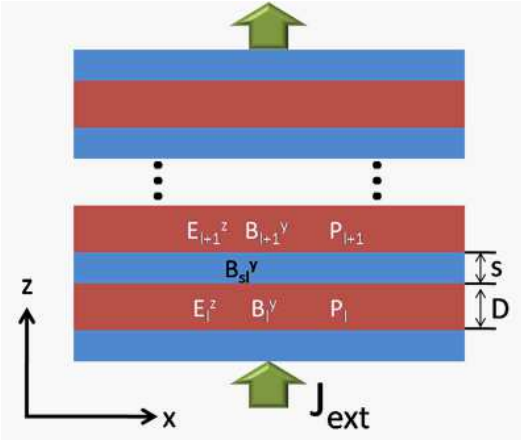,width=\columnwidth} \caption{\label{f1}(Color online). Schematic view of a stack of Josephson junctions based on the SIS model. The blue(pink) area denotes
superconducting(insulating) layers.}
\end{figure}

The geometry we consider is depicted in Fig. \ref{f1}. We neglect the thermal fluctuations so that the amplitude $|\Delta|$ of the superconducting order parameter
$|\Delta|\exp(i\psi)$ is constant. Furthermore, $\psi$ along the y axis is assumed to be uniform, namely we concentrate on the zero mode along this direction. The system is then
reduced to two dimensions. The density of energy stored in the superconducting layers, which consists of supercurrent energy and magnetic energy, then can be written as
\begin{equation}\label{eq1}
H_{sl}(x)=\frac{1}{8\pi}\int_{l(s+D)+D}^{(l+1)(s+D)}[\lambda_{s}^2(\mathbf{\nabla}\times B_{sl}^y)_{x}^2+{{B}_{sl}^y}^2]dz,
\end{equation}
where $\lambda_{s}$ is the penetration depth, and $B_{sl}^y$ is the magnetic field in the $l$th superconducting layer. The energy stored in the insulating layers is the sum of
energy of electromagnetic wave and Josephson energy
\begin{equation}\label{eq2}
H_{bl}(x)=\int_{l(s+D)}^{l(s+D)+D}[\frac{{B_l^y}^2}{8\pi}+\frac{\epsilon_c {E_l^z}^2}{8\pi}]dz+\frac{\Phi_0}{2\pi c}J_c(1-\cos P_l),
\end{equation}
where $B_l^y$ ($E_l^z$) is the magnetic (electric) field in the $l$th insulating layer (their variation along the $z$ direction in the $l$th insulating layer will be neglected in
later treatment). $J_{\rm{c}}$ is the critical current density, $\epsilon_{\rm{c}}$ dielectric constant along the $z$ axis and $c$ the light velocity in vacuum. $P_l$ is the gauge
invariant phase difference defined as
\begin{equation}{\label{eqPhase}}
P_l(x)=\psi_{l+1}(x)-\psi_l(x)-\frac{2\pi}{\Phi_0}\int_{l(s+D)}^{l(s+D)+D}A_l^z(x)dz,
\end{equation}
where $\Phi_0\equiv hc/2e$ is the flux quantum and $A_l^z$ is the vector potential. The magnetic field inside the superconducting layer can be evaluated from the London equation
$B_{sl}^y(z) = (\sinh [(s - z)/\lambda _{s} ]B_l^y  + \sinh [z/\lambda _{s} ]B_{l + 1}^y )/\sinh [s/\lambda _{s} ]$. In high-$T_c$ superconductors, the thickness of
superconducting layer $s$ and insulating layer $D$ is much smaller than $\lambda_{s}$, we have $B_{sl}^y\simeq[(s-z)B_l^y+zB_{l+1}^y]/s$. Then the total energy density can be
expressed as
\begin{widetext}
\begin{equation}\label{eq3}
H(x)=\sum_l[H_{sl}(x)+H_{bl}(x)]=\sum_{l}\{\frac{D}{8\pi}[(2\zeta+1){B_l^y}^2-\zeta(B_{l+1}^yB_l^y+B_l^yB_{l-1}^y)]+D{\frac{\epsilon_c {E_l^z}^2}{8\pi}+\frac{\Phi_0}{2\pi
c}J_c(1-\cos P_l)}\},
\end{equation}
\end{widetext}
where we have neglected the surface effect along the stack direction, which is valid for thick stacks of junctions. $\zeta\equiv\lambda_{s}^2/sD$ is the strength of inductive
coupling. It should be noted that the inductive coupling is very strong in BSCCO, see Table \ref{tbl1}.

\begin{table*}[t]\centering
\caption{Conversion of quantities among dimensionless, Gaussian and SI units. Here $\lambda_c$ and $\lambda_{ab}$ are
the penetration depth; $\epsilon_c$ is dielectric constant along the $z$ axis; $c$ is the light velocity in vacuum;
$\epsilon_0$ is dielectric constant in vacuum; $\omega_p=c/\lambda_c\sqrt{\epsilon_c}$ is the Josephson plasma
frequency. $J_c=c\Phi_0/8\pi^2\lambda_c^2D$ is the critical current density. In the present paper, we use
$\lambda_{c}=200\rm{\mu m}$, $\lambda_{\rm{ab}}=0.4\rm{\mu m}$, $\epsilon_c=10$, $s=0.3\rm{nm}$ and $D=1.2\rm{nm}$,
which are typical for BSCCO. Then $\lambda_s=\sqrt{sD/(s+D)^2}\lambda_{ab}=0.16\rm{\mu m}$. The two dimensionless
parameters are then $\beta=0.02$ and $\zeta=7.1\times10^4$. Following Ref. \cite{szlin08b}, we use slightly larger
$\zeta=4/9\times10^6$ in the present paper, and the results are insensitive to $\zeta$ as far as it is large. The
length of junction is $L=80{\rm{\mu m}}$.}
\begin{tabular}{ c | c | c | c | c | c | c | c | c | c}\hline\hline
 & length & time & conductance & electric field & voltage & magnetic field & Poynting vector & current & impedance \\
 \hline
Dimensionless &  $x$ & $t$ & $\beta$ & $E$ & $V$ & $B$ & $S$ & $J$ & $Z$\\[1ex]
Gaussian  &  $\lambda_cx$ & $t/\omega_p$ & $c\sqrt{\epsilon_c}\beta/4\pi\lambda_c$ & $\Phi_0\omega_pE/2\pi cD$ & $\Phi_0\omega_pV/2\pi c$ & $\Phi_0B/2\pi\lambda_cD$& $\Phi_0^2\omega_pEB/16\pi^3D^2\lambda_c$ &$\frac{Jc\Phi_0}{8\pi^2\lambda_c^2D}$ & $Z/\sqrt{\epsilon_c}$\\[2ex]
SI &  $\lambda_cx$ & $t/\omega_p$ & $c\sqrt{\epsilon_c\epsilon_0}\beta/\lambda_c$ & $\Phi_0\omega_pE/2\pi D$ & $\Phi_0\omega_pV/2\pi$ & $\Phi_0B/2\pi\lambda_c D$ & $\Phi_0^2\omega_pEB/4\pi^2\mu_0D^2\lambda_c$ &$\frac{J\Phi_0}{2\pi\lambda_c^2D\mu_0}$ & $Z/\sqrt{4\pi\epsilon_c\epsilon_0}$\\[1ex]
 \hline \hline
\end{tabular}
\label{tbl1}
\end{table*}

To find the relation between the magnetic field $B_l^y$ and $P_l$, we derive both sides of Eq. ({\ref{eqPhase}}) with
respect to $x$. With the London equation and Maxwell equation, we arrive at
\begin{equation}\label{eq0}
 \frac{\Phi_0}{2\pi D}\partial_x\mathbf{P}=\mathbf{M}\mathbf{B}^y,
\end{equation}
where $\mathbf{P}$ is a column vector $\mathbf{P}^T=[P_1, P_2, ..., P_N]$ with $N$ being the number of junctions. The column vectors for other quantities are defined in the same
way. $\mathbf{M}$ is the inductive coupling matrix defined as\cite{Sakai93,Krasnov97}
\begin{equation}\label{eqM}
\mathbf{M}   = \left[ {\begin{array}{*{20}c}
   {2\zeta+1} & {-\zeta} & 0 &  \cdots  & {} & {0}  & {-\zeta}  \\
   {-\zeta} & {2\zeta+1} & {-\zeta} & 0 &  \cdots   & {} & 0  \\
   0 &  \ddots  &  \ddots  &  \ddots  & 0 &  \cdots  & {}   \\
    \cdots  & 0 & { -\zeta} & {2\zeta+1} & {-\zeta} & 0 &  \cdots  \\
   {} &  \cdots  & 0 &  \ddots  &  \ddots  &  \ddots  & 0\\
   0 & {} &  \cdots  & 0 & { -\zeta} & {2\zeta+ 1} & { -\zeta}  \\
   { - \zeta} & 0 & {}  &  \cdots  & 0 & { -\zeta} & {2\zeta+ 1}  \\
\end{array}} \right],
\end{equation}
where the periodic boundary condition along the $z$ axis is imposed. Using the ac Josephson relation $\partial_t\mathbf{P}=2e\mathbf{E}^zD/\hbar$ and Eq. (\ref{eq0}), we can
rewrite the total energy density in a compact form
\begin{equation}\label{eq4}
H(x)=\frac{1}{2}\partial_x\mathbf{P}^T\mathbf{M}^{-1}\partial_x\mathbf{P}+\frac{1}{2}\partial_t\mathbf{P}^T\partial_t\mathbf{P}+\sum_l(1-\cos P_l),
\end{equation}
where the dimensionless quantities have been used, which, with the conversion among SI units and Gaussian units, are compiled in Table \ref{tbl1}. The first term at the right-hand
side of Eq. (\ref{eq4}) represents the magnetic energy, the second term the electric energy and the last term the Josephson coupling. The Lagrangian corresponding to Eq.
(\ref{eq4}) is
\begin{equation}\label{eq4b}
\mathscr{L}(x)=\frac{1}{2}\partial_x\mathbf{P}^T\mathbf{M}^{-1}\partial_x\mathbf{P}+\frac{1}{2}\partial_t\mathbf{P}^T\partial_t\mathbf{P}-\sum_l(1-\cos P_l).
\end{equation}
With the Euler-Lagrangian formula, we arrive at the coupled sine-Gordon equation
\begin{equation}\label{eq5}
\partial_x^2\mathbf{P}=\mathbf{M}[\sin \mathbf{P}+\partial_t^2\mathbf{P}],
\end{equation}
where $\sin \mathbf{P}\equiv[\sin P_1, \sin P_2, ..., \sin P_N]^T$. We consider the overlap geometry\cite{Sakai93} where the current is uniformly injected into the system. Taking
the dissipation and external current into account, we obtain the inductively coupled perturbed sine-Gordon equation
\begin{equation}\label{eq5a}
\partial_x^2\mathbf{P}=\mathbf{M}[\sin \mathbf{P}+\partial_t^2\mathbf{P}+\beta\partial_t\mathbf{P}-\mathbf{J}_{\rm{ext}}],
\end{equation}
where the first term at the right-hand side of Eq. (\ref{eq5a}) is the Josephson current, the second term the
displacement current, the third term the quasiparticles contribution and the last term the external current. Writing
down the equation for $P_l$ in Eq. (\ref{eq5a}) explicitly, we have
\begin{equation}\label{eq5b}
\partial_x^2P_l=(1-\zeta\Delta^{(2)})[\sin P_l+\partial_t^2P_l+\beta\partial_tP_l-J_{\rm{ext}}],
\end{equation}
where $\Delta^{(2)}$ is the finite difference operator defined as $\Delta^{(2)}f_l\equiv f_{l+1}+f_{l-1}-2f_l$. Besides the inductive coupling in Eq. (\ref{eq5a}), a capacitive
coupling\cite{Koyama96, Machida99} and a coupling originating from non-equilibrium effects\cite{Ryndyk98, Rother03} are also present in intrinsic Josephson junctions. These two
couplings are weak in comparison to the inductive coupling and are neglected in the present work.

The above calculations are based on the SIS model (superconductor-insulator-superconductor), which is a good model for
artificially stacked Josephson junctions. However, for BSCCO, the superconducting layer is only of atomic thickness,
and thus the quantity $\lambda_s$ is not well defined. To find the relation between $\lambda_s$ and the measurable
penetration depth $\lambda_{ab}$, we need to resort to the Lawrence-Doniach model, which has already been discussed
extensively in literatures. The connection between $\lambda_s$ and $\lambda_{ab}$ is given by
$\lambda_{s}=\sqrt{sD/(s+D)^2}\lambda_{ab}$.\cite{Bulaevskii96,Koshelev01}

In the presence of dissipations and a power input, the energy oscillates with time according to
\begin{equation}\label{eq6}
\partial_tH(x)=\partial_{xt}\mathbf{P}^T\mathbf{M}^{-1}\partial_x\mathbf{P}+(\partial_t^2\mathbf{P}^T+\sin
\mathbf{P}^T)\partial_t\mathbf{P}.
\end{equation}
With the help of Eq. (\ref{eq5a}), we have for the steady state
\begin{widetext}
\begin{equation}\label{eq7}
\int_0^T\int_0^L {\partial _t Hdxdt}  = \int_0^T {(\partial _t
\mathbf{P}^T \mathbf{M}^{ - 1}
\partial _x \mathbf{P})|_0^L } dt + \int_0^T\int_0^L {\partial _t \mathbf{P}^T \mathbf{J}_{\rm{ext}} dxdt}  -
\beta \int_0^T\int_0^L {\partial _t \mathbf{P}^T \partial _t
\mathbf{P}dxdt}=0,
\end{equation}
\end{widetext}
where $L$ is the length of junctions and $T$ is the period of plasma oscillation. Rewriting Eq. (\ref{eq7}) in a more transparent form, we have the power balance condition
\begin{equation}\label{eq8}
\int_0^T {(\mathbf{E}^T \mathbf{B})|_0^L } dt +
LT\mathbf{E}_{\rm{dc}} ^T \mathbf{J}_{\rm{ext}} - \beta
\int_0^T\int_0^L {\mathbf{E}^T \mathbf{E}dxdt} = 0,
\end{equation}
where the first term is the power gain (loss) at edges due to irradiation (radiation). The second term is the input power with $\mathbf{E}_{\rm{dc}}$ the dc electric field, and
the last term is the energy loss due to dissipations. It should be remarked that Eq. (\ref{eq8}) is general and should be valid at different states. As will be shown later, the
power balance relation is useful when the plasma oscillation is strong and the linear expansion fails. From Eq. (\ref{eq8}) and
$\mathbf{E}=\mathbf{E}_{\rm{ac}}+\mathbf{E}_{\rm{dc}}$, we can see that when the oscillation in electric field $\mathbf{E}_{\rm{ac}}$ is small, the \emph{IV} curve is almost ohmic
$\mathbf{E}_{\rm{dc}}\approx \mathbf{J}_{\rm{ext}}/\beta$. To have strong radiation, the oscillation of electric field in the junctions should be large. Therefore the optimal
state for the radiation is a state having nonlinear \emph{IV} characteristics where a large part of the input power can be pumped into plasma oscillation. The problem then boils
down to finding such a state with highly nonlinear \emph{IV} characteristics.

\begin{figure}[b]
\psfig{figure=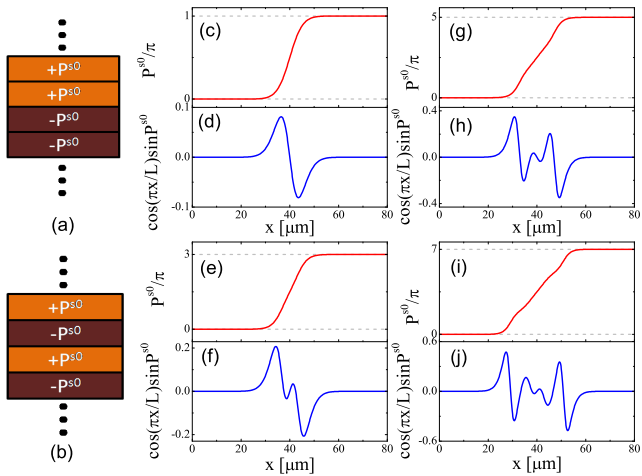,width=\columnwidth} \caption{\label{f4}(Color online). (a), (b): Schematic view of two simplest configurations of the static phase $P_l^s(x)$. (c), (e), (g)
and (i): $(2m+1)\pi$ phase kink for $m=0, 1, 2$ and $3$ respectively. (d), (f), (h) and (j): their corresponding un-quantized static vortices.}
\end{figure}

The relation between the oscillating magnetic field and electric field at the edges of junctions is given by the
boundary condition. The boundary condition depends on many effects such as distribution of the order parameter near
edges, the geometry of the sample and the dielectric materials attached to the sample. In Refs.
\cite{Bulaevskii06,Bulaevskii06PRL}, the dynamic boundary condition is derived from the electromagnetic wave equations
inside the dielectrics. In the present paper, we use an effective impedance as the boundary condition
\begin{equation}\label{eq8b}
\mathbf{E}_{\rm{ac}}/\mathbf{B}_{\rm{ac}}=Z=|Z(\omega)|\exp(i\theta(\omega)),
\end{equation}
where $|Z|$ and $\theta$ are parameters, and $\omega$ is the frequency. This boundary condition is general and any
other boundary condition can be casted into this form. It has been pointed out that there is a significant impedance
mismatch between the intrinsic Josephson junctions and outside space because of the small ratio between the thickness
of the stack and the penetration depth $\lambda_c$.\cite{Koshelev08} This means $|Z|>>1$, which is similar to the
single junction case.\cite{Langenberg65}

The radiation power counted by Poynting vector at one edge with an
effective impedance becomes
\begin{equation}\label{eqP9}
S_r=\frac{1}{2TN}\int_0^T{\rm{Re}}[{\mathbf{E}_{\rm{ac}}^{\dag}}\mathbf{B}_{\rm{ac}}]dt
=\frac{\cos\theta}{2T|Z|N}\int_0^T{\mathbf{E}_{\rm{ac}}^\dag}\mathbf{E}_{\rm{ac}}dt.
\end{equation}
For ease of theoretical calculation, we consider the situation that the radiation does not substantially change the plasma oscillation inside the junctions. In this case,
$\mathbf{E}_{\rm{ac}}$ can be evaluated \emph{without} radiation, i.e. with the simple boundary condition $\partial_x\mathbf{P}=0$. This treatment is valid when the impedance
mismatch is significant, which will be shown later to be rather accurate by numerical simulations. When the impedance $|Z|$ is small, however, one should take the effect of
radiation into account in the calculation of the plasma dynamics self-consistently.

\section{Solutions} In this section, we first construct a general procedure to solve the coupled sine-Gordon equation from the spectrum analysis. There exist longitudinal and
transverse plasma modes in a stack of junctions. As the stack itself forms a cavity, the plasma component can be written as $\widetilde{P}_l(x)\sim \cos (k_jx) \sin(
ql/(N+1))$\cite{Kleiner94,sakai94} with $k_j=j\pi/L$, and $q$, $j$ integers. There are $N$ different dispersion branches with characteristic velocity
$c_q=1/\sqrt{1+2\zeta[1-\cos(q\pi/(N+1))]}$. When the stack is thick enough, the plasma oscillation uniform along the $c$ axis becomes possible and its velocity is $c_0=1$. We
will concentrate on this case in the present work since it supports the strong radiation. Without an external in-plane magnetic field, the solution including all frequency
harmonics subject to the boundary condition $\partial_xP_l=0$ can be expressed as\cite{szlin08b}
\begin{equation}\label{tEq1}
P_l(x,t)=\omega t+P_l^s(x)+\sum_{j=1}^{\infty}{\rm{Re}}[-i A_j\exp(ij\omega t)]\cos(k_jx),
\end{equation}
where $\omega t$ is the rotating part, $P_l^s$ the static phase kink and the last term is the plasma oscillation
including all harmonics, with $A_j$ the oscillation amplitude. For simplicity, the first cavity mode along the $x$ axis
with $k_1=\pi/L$ is considered and the small time dependence of $P_l^s$ is neglected.\cite{Koshelev08b,szlin08b}
Putting the $m$th frequency component of the Josephson current $\sin P_l$ as $S_m\exp(im\omega t)$, and expanding sine
of sine with Bessel functions, we have
\begin{equation}\label{tEq1a}
\left\{ {\begin{array}{*{20}c}
   {S_m^l  = G_m^l + {G_{-m}^{l*}}, \rm{\ \ \ \ \ \ \ \      for\ }\emph{m} \ge 1},  \\
   {S_0^l  = G_0^l, \rm{\ \ \ \ \ \ \ \ \ \ \ \ \ \ \ \ \ \ \ \      for\ }\emph{m} = 0},  \\
\end{array}} \right.
\end{equation}
where

\begin{widetext}
\begin{equation}\label{tEq1b}
G_m^l(x) = - i\sum\limits_{\{q_j=-\infty \}}^{+\infty} \delta (\sum\limits_{j = 1}^{\infty } {q_j j} ,m - 1)\left[ {\prod\limits_{j = 1}^{\infty } {J_{q_j } (|A_j |\cos(k_j x))} }
\right]\exp[i(P_l^s+\sum\limits_{j = 1}^{\infty} {q_j\phi_j})],
\end{equation}
with $J_{q_j}$ the Bessel function of the first kind. $\sum\limits_{\{q_j=-\infty \}}^{+\infty}$ is the summation over the ensemble of $q_j$'s, and $A_j=|A_j|\exp(i\phi_j)$.
Substituting Eq. (\ref{tEq1}) into the coupled sine-Gordon Eq. (\ref{eq5b}) and comparing each frequency component in $\sin P_l$, we have for the $m$th ($m\ge1$) component,
\begin{equation}\label{tEq3}
[ik_m^2-i(m\omega)^2-\beta
m\omega]A_m\cos(k_mx)=(1-\zeta\Delta^{(2)})S_m^l.
\end{equation}
\end{widetext}

From Eq. (\ref{tEq3}), $A_m$ is given by
\begin{equation}\label{tEq5}
A_m=\frac{F_m}{ik_m^2-i(m\omega)^2-m\beta\omega},
\end{equation}
with
\begin{equation}\label{tEq6}
F_m\equiv\frac{2}{L}\int_0^L(1-\zeta\Delta^{(2)})S_m^l\cos(k_mx)dx.
\end{equation}
The functional $F_m$ represents the coupling of the plasma to the cavity modes, which is the central quantity for the excitation of plasma. Other mechanism such as the modulation
of the critical current is also possible\cite{Matisoo69,Koshelev08}, although it is practically very hard to achieve a homogeneous modulation along the $z$ axis. In the present
solution, $P_l^s$ is inherently one part of the solution. Since we have assumed that phase is uniform along the $y$ direction, or equivalently, we have considered the $(1,0)$
cavity mode, only the kink in the $x$ direction contributes to $F_m$. If there exits a kink along the $y$ direction simultaneously, the functional $F_m$ will be enhanced
further.\cite{Hu08} In the case of cylinder geometry, the plasma is coupled to the cavity fully via the kink so that $F_m$ is maximized.\cite{Hu08}

It should be noted that $A_m$ is independent of $l$ in Eq. (\ref{tEq5}), which imposes a constraint on the arrangement of $P_l^s$ in the $z$ direction. As will be shown later,
periodic arrangements such as those in Figs. \ref{f4}(a) and (b) diagonalize the finite difference operator and make $F_m$ independent of $l$.

For the $0$th component (static part), we obtain
\begin{equation}\label{tEq4}
\partial_x^2P_l^s-\beta\omega+J_{\rm{ext}}=(1-\zeta\Delta^{(2)})S_0^l.
\end{equation}
The current conservation relation reads $J_{\rm{ext}}=\beta\omega+\langle S_0^l\rangle_x$ ($\langle...\rangle_x$ is the spatial average). The remaining terms in Eq. (\ref{tEq4})
which do not contribute to the net current is
\begin{equation}\label{tEq7}
\partial_x^2P_l^s=-\zeta\Delta^{(2)}S_0^l,
\end{equation}
where $\zeta>>1$ is taken into account.

From Eqs. (\ref{tEq1}), (\ref{tEq5}) and (\ref{tEq7}), we can calculate the plasma oscillation, \emph{IV} characteristics and $P_l^s$. We consider explicitly the case where the
fundamental mode $A_1$ is small and thus higher harmonics can be safely neglected. We also approximate $J_1(|A_1|\cos(k_1x))\simeq|A_1|\cos(k_1x)/2$ and
$J_0(|A_1|\cos(k_1x))\simeq1$. We will refer to this approximation as linear approximation in later discussions, and the validity of this approximation will become clear later. It
should be noted that the nonlinearity of the coupled sine-Gordon equation is still retained in the equation for $P_l^s$, Eq. (\ref{tEq7}). With this approximation, we can
calculate $A_1$
\begin{equation}\label{tEq8}
A_1=\frac{F_1}{ik_1^2-i\omega^2-\beta\omega},
\end{equation}
where
$F_1=\frac{-2i}{L}\int_0^L(1-\zeta\Delta^{(2)})\exp(iP_l^s)\cos(k_1x)dx$.

The \emph{IV} characteristic is given by the current conservation
\begin{equation}\label{tEq9}
J_{\rm{ext}}=\beta\omega+\langle
S_0^l\rangle_{x}=\beta\omega+\frac{\beta\omega|F_1|^2/4}{(k_1^2-\omega^2)^2+\beta^2\omega^2},
\end{equation}
where the first term at the right-hand side is the normal current and the second term is the dc part of the Josephson
current.

The equation for $P_l^s$ is given by
\begin{equation}\label{tEq10}
\partial_x^2P_l^s=\frac{A_1\zeta i}{2}\cos(k_1x)\Delta^{(2)}\exp(-iP_l^s).
\end{equation}
Equation (\ref{tEq10}) has many solutions, such as the trivial vacua solution and solutions with phase kink, which will be discussed separately in the following subsections.

\begin{figure}[b]
\psfig{figure=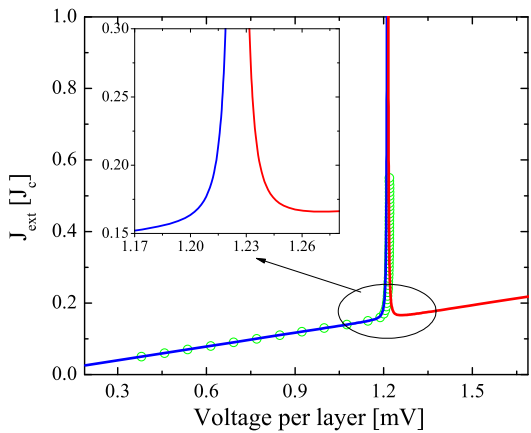,width=\columnwidth} \caption{\label{f5}(Color online). \emph{IV} characteristics calculated by numerical simulations (symbols) and the linearized theory
(lines). The inset is an enlarged view.}
\end{figure}

\begin{figure}[t]
\psfig{figure=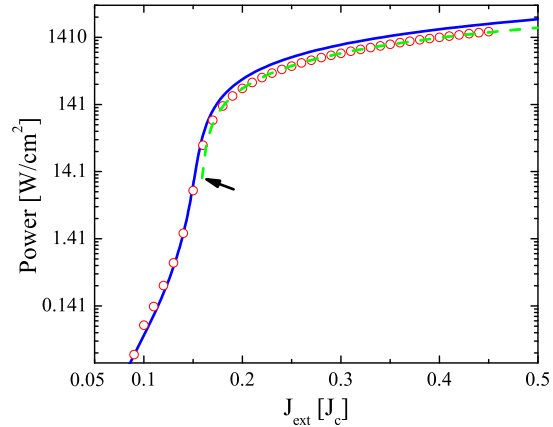,width=\columnwidth} \caption{\label{f4b}(Color online). (a) Radiation power at the first current step obtained by simulations (symbol), the linear theory
(solid line) and the power balance relation Eq. (\ref{tEq15}) (dashed line). The results are obtained with $C=0.000177$ and $R=707.1$. The arrow is the starting point of the first
current step.}
\end{figure}
\subsection{State with kink}
Equation (\ref{tEq10}) has solutions with $(2m+1)\pi$ kink with $m$ being an integer.\cite{szlin08b,Koshelev08b} Let us consider the two simplest periodic configurations of
$P_l^s$ depicted in Figs. \ref{f4}(a) and (b) where $P_l^s=f_lP^{s0}$ with $f_l=\pm1$ depending on $l$, which diagonalize Eq. (\ref{tEq10})
\begin{equation}\label{tEq13}
\partial_x^2P^{s0}=\zeta q{\rm{Re}}(A_1)\cos(k_1x)\sin P^{s0},
\end{equation}
where $q=1$ for the configuration in Fig. \ref{f4}(a) and $q=2$ for the configuration in Fig. \ref{f4}(b). It should be
noted that other periodic configurations are also possible. Equation (\ref{tEq13}) is invariant under the
transformation $x\leftarrow L-x$ and $P^{s0}\leftarrow (2m+1)\pi-P^{s0}$, which clearly renders a kink at the center of
junction. Equation (\ref{tEq13}) subject to the boundary condition $\partial_xP^{s0}=0$ is solved numerically and the
results are detailed in the Fig. \ref{f4}, where the $(2m+1)\pi$ phase kink of characteristic length
$\lambda_P\equiv1/\sqrt{\zeta q |{\rm{Re}}(A_1)|}$ is at the center of junction $x=L/2$. It is this $(2m+1)\pi$ phase
kink that pumps the dc power into plasma oscillation.

In Fig. \ref{f4}, we can see that $\partial_x^2P^{s0}$ forms an unquantized static vortex with characteristic length $\lambda_P$, therefore it doesn't contribute to the net
supercurrent $\langle\sin P_l\rangle_{xt}$. There is a dc magnetic field in each layer associated with the static vortex. As it points in opposite directions in different
junctions, the total magnetic field across the intrinsic Josephson junctions vanishes. Therefore it is impossible to realize the kink state in a single junction.

In addition to the $(2m+1)\pi$ phase kink, there exist the well known solitons with $2\pi$ phase variation superposing
to the $(2m+1)\pi$ phase kink, as observed in our simulations (not shown in Fig. \ref{f4}). In the region of
$|x_0-L/2|>>\lambda_P$, because $\cos(k_1x)$ is almost a constant in the narrow region $\lambda'\equiv1/\sqrt{\zeta q
|{\rm{Re}}(A_1)\cos(k_1x_0)|}$, Eq. (\ref{tEq13}) can be approximated as
\begin{equation}\label{tEq13aa}
\lambda'^2\partial_x^2P^{s0}=\sin P^{s0}.
\end{equation}
Equation (\ref{tEq13aa}) has the usual soliton solution $P^{s0}=4\arctan[\exp((x-x_0)/\lambda')]$. The total resultant
$P_l^s$ is $(2m+1)\pi$ phase kink at the center of junction and solitons with $\pm2\pi$ phase variation away from the
center of junction. The solitons with $\pm2\pi$ phase variation don't contribute to $F_1$ as well as the net
supercurrent, and therefore are omitted in the following discussions.

\begin{figure}[b]
\psfig{figure=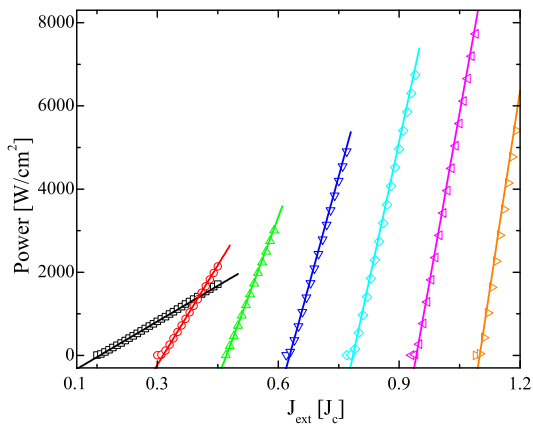,width=\columnwidth} \caption{\label{f4c}(Color online). Radiation power on current steps. The symbols are from the simulations and the lines are given by
Eq. (\ref{tEq15}). The results are obtained with $C=0.000177$ and $R=707.1$.}
\end{figure}

\begin{figure*}[t]
\psfig{figure=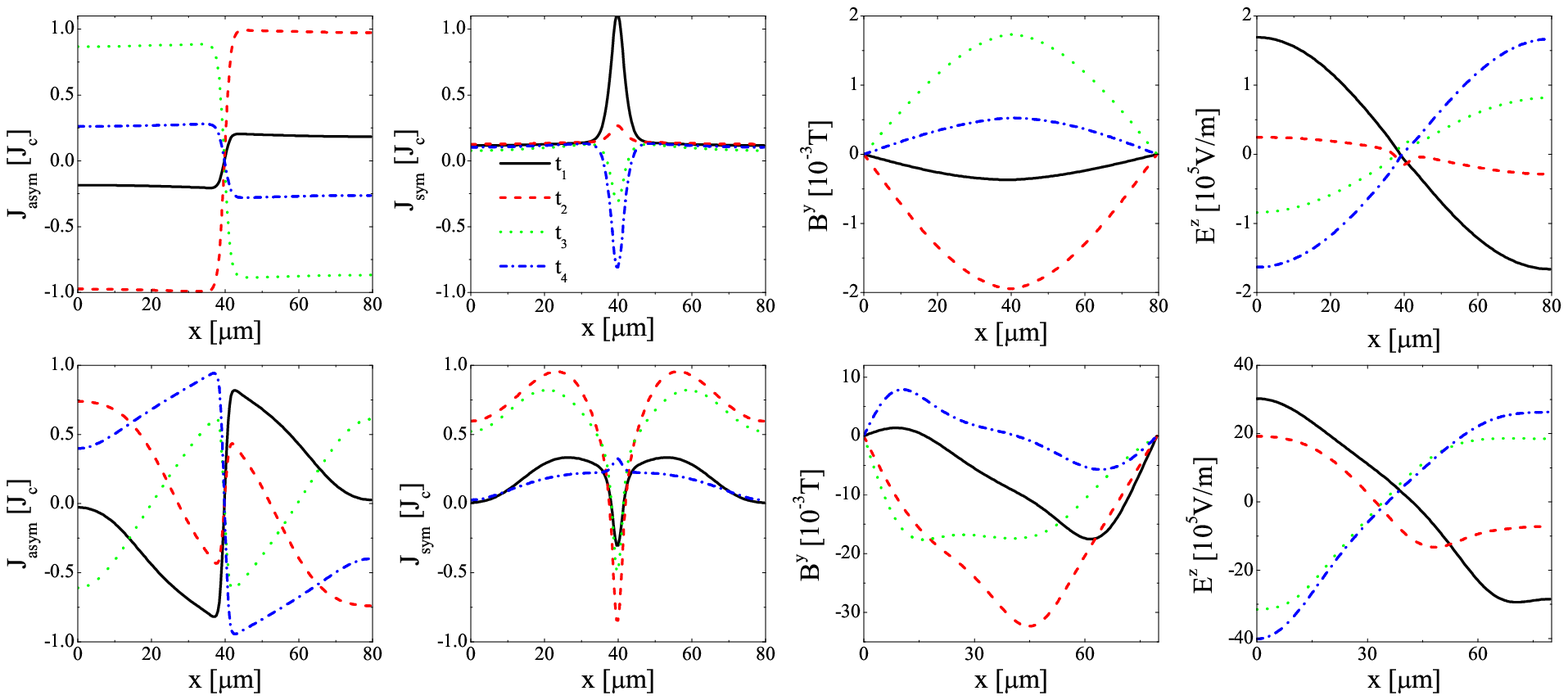,width=18cm} \caption{\label{f4d}(Color online). Configurations of $J_{\rm{asym}}$, $J_{\rm{sym}}$ and electromagnetic wave at successive time
$t_1=0.06T_1$, $t_2=0.31T_1$, $t_3=0.83T_1$ and $t_4=0.94T_1$, where $T_1\equiv 2\pi/k_1$ is the period of plasma oscillation at the first cavity mode. We have subtracted the
static part of $E^z$. The top figures are taken at the bottom of the first current step while the bottom figures are at the top of the first current step in Fig. \ref{f5}.}
\end{figure*}

The \emph{IV} characteristics shown in Fig. \ref{f5} is calculated from Eq. (\ref{tEq9}). One remarkable feature in the \emph{IV} characteristics is the self-induced current step,
i.e., \emph{IV} branch with constant voltage. From Eq. (\ref{tEq9}), it is found that the \emph{IV} characteristics for kink solutions with different $m$, e.g., the kink solutions
in Fig. \ref{f4} (c),(e),(g) and (i), is almost the same, because the kink renders itself approximately as a step function and the dc current contributed from the kink region of
width $\lambda_P$ is negligible. It should be noted that there exist two branches at the cavity resonance, and the right branch has a negative differential resistance.

The radiation power can be readily calculated from Eq. (\ref{eqP9}). With the linear approximation, the power is
\begin{equation}\label{tEq13a}
S_{\rm{r}}=\cos\theta |A_1\omega|^2/2|Z|.
\end{equation}
The dependence of $S_{\rm{r}}$ on $\theta$ and $|Z|$ is consistent with the results shown in Fig. 5 of Ref.
\cite{szlin08b}. The results of $S_{\rm{r}}$ are displayed in Fig. \ref{f4b}. Similar to the \emph{IV} characteristics,
the radiation powers for the states with different phase-kink configurations are almost the same.

To check the applicability of the analytical treatment, we solve the equation of motion Eq. (\ref{eq5b}) by computer simulations.\cite{szlin08b} The time step in all simulations
is set to $\Delta t=0.0018$ and the mesh size is set to $\Delta x=0.002$. The accuracy is checked with smaller $\Delta t$ and $\Delta x$. We use the periodic boundary condition
along the $z$ axis to minimize the surface effect, and attach an effective RC circuit to the junctions as the boundary condition along the $x$
direction.\cite{Gronbechjensen89,Soriano96} In this case, the impedance is $Z=R-i/C\omega$, where $R$ is the resistance and $C$ is the capacitance of the RC circuit. $R$ and $C$
are chosen to make sure that $|Z|>>1$.

The simulation results of the \emph{IV} characteristics and radiation power are presented in Figs. \ref{f5} and
\ref{f4b}. For the \emph{IV} characteristics, there is a good agreement between the theory and simulation, except that
the theory is incapable of describing the height of the current step. For the radiation power, the linear theory is
valid off/near resonance but fails inside the current steps. The failure of the analytical treatment is caused by the
strong plasma oscillation and existence of harmonics in the current step.\cite{szlin08b}

To derive a better estimate of the power radiation at the current steps, we resort to the power balance equation which is valid in the whole region of the \emph{IV}
characteristics. Taking the plasma solution Eq. (\ref{tEq1}) and substituting into the power balance equation Eq. (\ref{eq8}), we obtain the \emph{IV} characteristics in the
presence of radiation
\begin{equation}\label{tEq14}
\omega J_{\rm{ext}}  = \beta \omega ^2  + \frac{\beta}{4} \sum\limits_{j = 1}^{\infty } {(j\omega A_j )^2 } + \frac{{\cos \theta }}{{L|Z|}}\sum\limits_{j = 1}^{\infty } {(j\omega
A_j )^2 }.
\end{equation}

In other words, the radiation power can be evaluated if we know the \emph{IV} characteristics. Therefore it is useful
to introduce an effective conductance $\beta'$ defined as
\begin{equation}\label{tEq14a}
J_{\rm{ext}}=\beta'\omega=(\beta+\beta_{\rm{d}}+\beta_{\rm{r}})\omega,
\end{equation}
where $\beta_{\rm{d}}\equiv\frac{\beta}{4} \sum\limits_{j=1}^{\infty} {(jA_j)^2}$ is the conductance due to damping of plasma oscillation,
$\beta_{\rm{r}}\equiv\frac{{\cos\theta}}{{L|Z|}}\sum\limits_{j = 1}^{\infty}{(jA_j)^2}$ the conductance due to radiation at both edges. From Eq. (\ref{tEq14}), we can calculate
$A_j$ from the \emph{IV} characteristics. The radiation power at one edge then can be evaluated by
\begin{equation}\label{tEq15}
S_{\rm{r}}=\beta_{\rm{r}}\omega^2/2= J_e \omega /(\frac{{\beta
|Z|}}{{2\cos \theta }} + \frac{2}{L}),
\end{equation}
where $J_{\rm{e}}\equiv J_{\rm{ext}}-\beta\omega$ is the excess current. From the foregoing analysis, if we can screen the radiation at one edge, the radiation at the other edge
is enhanced. It should be remarked that not the whole energy pumped into plasma oscillation radiates into outside space. Most part of it is damped by dissipations inside the
intrinsic Josephson junctions. The radiation power at current steps obtained by numerical simulations and power balance condition is depicted in Fig. \ref{f4c}. In contrast to the
linear approximation, the estimated radiation power by Eq. (\ref{tEq15}) is consistent with simulations even inside the current steps where the amplitude of plasma oscillation is
large and high harmonic components are present. The power increases linearly with the $J_{\rm{ext}}$ and the maximum power is as high as $8000\rm{W/cm^2}$ from simulations (at the
$6$th cavity mode). The maximal total radiation power at the first cavity mode is about $10$\rm{mW} if we use a mesa of similar dimension as the experiments\cite{Ozyuzer07}, which
is capable of practical applications.

The cavity quality factor $Q_c$ at the cavity resonance $\omega=k_1$ is given by
\begin{equation}\label{eq15a}
Q_c\equiv\omega\frac{\rm{Energy\ Stored}}{\rm{Power\
Loss}}=\frac{\omega}{\beta+4\cos\theta/L|Z|},
\end{equation}
which has the order of magnitude of $100$ for $\beta=0.02$ and $|Z|>>1$. The half-width of the radiation frequency spectrum $\Gamma=\omega/Q_c$ is about $10\rm{GHz}$, so that the
radiation is almost monochromatic. The efficiency defined as the ratio of the radiation power to the total power input is
\begin{equation}\label{eq16}
Q_e= \frac{J_{\rm{e}} }{J_{\rm{ext}}}/(\frac{\beta |Z|}{4\cos
\theta} + \frac{1}{L}).
\end{equation}
The efficiency $Q_e$ at the current step corresponding to a lower cavity mode is larger than that of a higher mode because of the smaller ohmic dissipations. $Q_e$ at the top of
the first current step in Fig. \ref{f4c} is about $7.5\%$.

The distribution of the $c$-axis uniform EM wave along the $x$ direction of the junctions, as well as the supercurrent, obtained from simulations at the bottom and top of the
first current step are shown in Fig. \ref{f4d}. The supercurrent has the same period as $P_l^s$ along the $c$ axis, and we only visualize it at one layer. We divide the Josephson
current into the symmetric part $J_{\rm{sym}}$ and antisymmetric part $J_{\rm{asym}}$ with respect to the center of junction, i.e., $\sin P_l(x)=J_{\rm{asym}}(x)+J_{\rm{sym}}(x)$.
Off resonance, $J_{\rm{sym}}$ is zero except the center of junction, while $J_{\rm{asym}}$ oscillates from $-1$ to $+1$; the magnetic field is symmetric and electric field is
antisymmetric with respect to the center of junction. However, at the top of the current step, the even part of Josephson current becomes more important, so the radiation is of
dipole type. The corresponding distribution for EM wave is neither symmetric nor antisymmetric because the higher harmonics in Eq. (\ref{tEq1}) become important.

\begin{figure}[b]
\psfig{figure=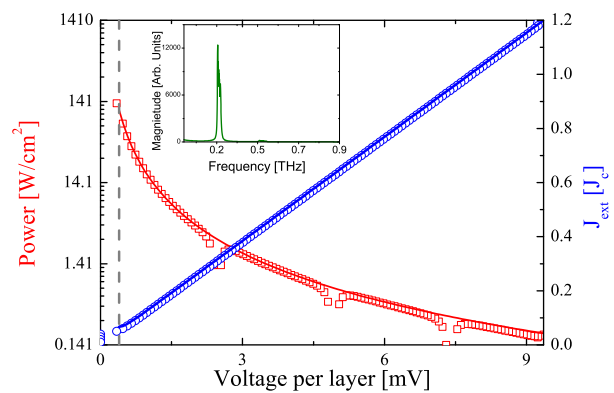,width=\columnwidth} \caption{\label{f3}(Color online). Radiation power from the state without
kink and its corresponding \emph{IV} characteristics. Symbols are for simulations and lines are for theory. The
vertical dashed line is the retrapping point obtained from the theory. The inset is the frequency spectrum at the
strongest radiation. The results are obtained with $C=0.716$ and $R=10.0$.}
\end{figure}

\begin{figure*}[t]
\psfig{figure=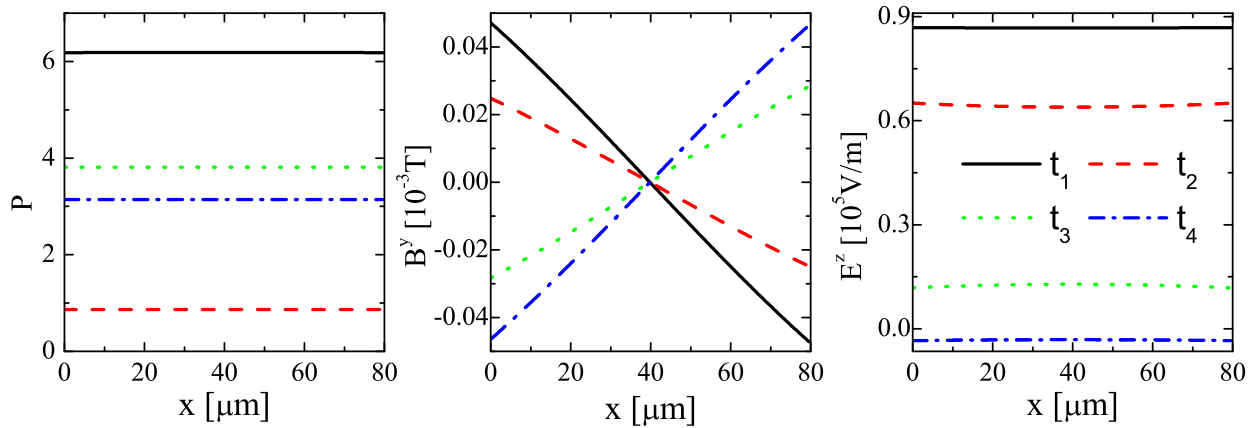,width=17cm} \caption{\label{f6}(Color online). Configurations of phase $P$, magnetic field $B^y$ and electric field $E^z$ at successive time
$t_1=0.07T_{\rm{r}}$, $t_2=0.17T_{\rm{r}}$, $t_3=0.33T_{\rm{r}}$ and $t_4=0.4T_{\rm{r}}$, where $T_{\rm{r}}$ is the period at the retrapping point. The phase is normalize into
$[0, 2\pi]$ and we have subtracted the static part of $E^z$. The results are obtained by simulations with $C=0.716$ and $R=10.0$.}
\end{figure*}

\subsection{State without kink}
Equation (\ref{tEq10}) also has trivial vacua solutions $P^{s0}=m\pi$. Without losing generality, we take $m=0$. From Eq. (\ref{tEq8}), we know that the transverse plasma cannot
exist without a kink. Therefore the solution becomes
\begin{equation}\label{tEq10aa}
P_l=\omega t-iA_1\exp(i\omega t),
\end{equation}
which is nothing but the McCumber solution. Here only the fundamental mode is taken, which is sufficient because of the small plasma oscillation in this solution. Then Eq.
(\ref{tEq8}) is reduced to
\begin{equation}\label{tEq10a}
A_1=1/(\omega^2-i\beta\omega),
\end{equation}
and its corresponding \emph{IV} characteristics without radiation is
\begin{equation}\label{tEq10b}
J_{\rm{ext}}=\beta\omega+\frac{\beta }{{2(\omega ^3  + \beta ^2
\omega )}}.
\end{equation}
The radiation power at one edge obtained with Eq. (\ref{eqP9}) is
\begin{equation}\label{tEq11a}
S_{\rm{r}}=\cos\theta/[2(\omega^2+\beta^2)|Z|].
\end{equation}
From the power balance condition, the \emph{IV} characteristics with radiation is given by
\begin{equation}\label{tEq12}
J_{\rm{ext}} \omega  = \beta \omega ^2  + \frac{\beta }{{2(\omega ^2
+ \beta ^2 )}} + \frac{{\cos \theta }}{{(\omega ^2  + \beta ^2
)|Z|L}},
\end{equation}
where the last term represents the correction due to radiation. The minimum value of $J_{\rm{ext}}$ in Eq.
(\ref{tEq12}) is the retrapping current $J_{\rm{r}}$, at which the input power becomes insufficient for the phase
particle to travel across the damped tilted washboard potential. In the weak damping limit $\beta<<1$ as in the present
system, $J_{\rm{r}}$ is
\begin{equation}\label{tEq12a}
J_{\rm{r}}  = \frac{4}{3}\beta ^{3/4} \left(\frac{{3\cos \theta}}{{|Z|L}} + \frac{{3\beta }}{2}\right)^{1/4}.
\end{equation}
Its corresponding voltage is $\omega_{\rm{r}}=(1.5+3\cos\theta/|Z|L\beta)^{1/4}>1$, which justifies the approximation made in Eqs. (\ref{tEq10aa}) and (\ref{tEq10a}).

The \emph{IV} characteristics calculated from the analytic formula and numerical simulations are presented in Fig. \ref{f3}. A good agreement between the simulation and theory can
be spotted. This further verifies the approximation of neglecting the effect of radiation on the phase dynamics in junctions when $Z$ is large. The radiation power increases
continuously with decreasing current and reaches the maximum at the retrapping point. The local minima in the curve are caused by the small spatial modulation of electromagnetic
field, which changes with the voltage, similar to a cavity behavior. The frequency harmonics is undiscernible even at the maximum radiation because the plasma oscillation is weak.
The distributions for $P$, $B^y$ and $E^z$ obtained by numerical simulations with open boundary condition are displayed in Fig. \ref{f6}, where there is a small phase gradient
created by radiation which is hard to see in the present scale. The magnetic field is antisymmetric with respect to the center of the junction, while the electric field is almost
uniform along the $x$ direction, except the small modulation created by radiation.

The state without kink (McCumber state) is unstable in the certain region of \emph{IV} curve. In a long Josephson
junction, the system evolves into soliton states due to the parametric instability.\cite{Pagano86} We have investigated
the stability of the state without kink in a stack of intrinsic Josephson junctions. The system favors the state with
kink due to the instability of the state without kink near the cavity resonance.
\begin{figure*}[t]
\psfig{figure=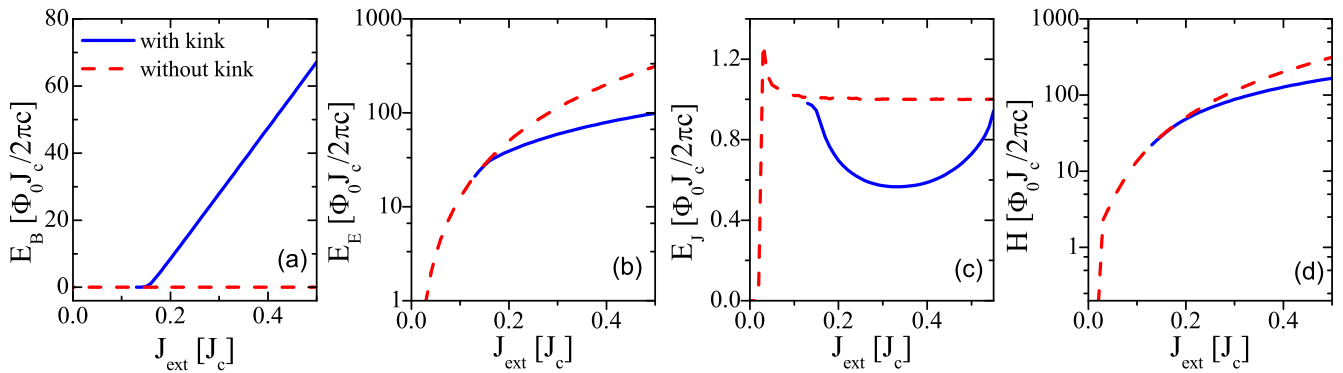,width=18cm} \caption{\label{f7}(Color online). Energy per junction stored in a thick stack of intrinsic Josephson junctions in the state with and without
kink, (a) magnetic, (b) electric, (c) Josephson, (d) total energy. The results are obtained without radiation, and the results for the state with kink are obtained in the first
current step shown in Fig. \ref{f5}. Because the energy for different kink configurations is same, only the energy for the kink configuration ${\mathbf{P}^s}^T=[-\pi, +\pi, -\pi,
+\pi]$ is shown in this figure.}
\end{figure*}

\section{Energetic analysis}
Similar to conventional laser systems, most part of the input power is stored and dissipated in the junctions and only a small portion radiates into space. Therefore it is worthy
of looking at the energy oscillation inside the junctions. One might consider that the state with kink costs more energy than that without kink. To see whether the state with kink
can be realized in reality, it is necessary to know the energy cost to construct the kink. In this section, we calculate the energy stored in the intrinsic Josephson junctions.

As can be read from Eq. (\ref{eq4}), the system energy consists of the magnetic energy $E_{\rm{B}}$, electric energy $E_{\rm{E}}$ and Josephson coupling $E_{\rm{J}}$,
\begin{eqnarray*}
  E_{\rm{B}} &=& \langle\partial_x\mathbf{P}^T\mathbf{M}^{-1}\partial_x\mathbf{P}\rangle_{xt}/2N, \\
  E_{\rm{E}} &=& \langle\partial_t\mathbf{P}^T\partial_t\mathbf{P}\rangle_{xt}/2N, \\
  E_{\rm{J}} &=& \langle\sum_l(1-\cos P_l)\rangle_{xt}/N,
\end{eqnarray*}
where the energy has been normalized by the number of layers $N$. In the state with kink, the magnetic energy has contribution from the static kink
$E_{\rm{Bs}}=\langle{\partial_x\mathbf{P}^s}^T\mathbf{M}^{-1}\partial_x\mathbf{P}^s\rangle_{xt}/2N$ and from the plasma oscillation
$E_{\rm{Bp}}=\langle{\partial_x\widetilde{\mathbf{P}}}^T\mathbf{M}^{-1}\partial_x\widetilde{\mathbf{P}}\rangle_{xt}/2N$. Here we show that $E_{\rm{Bs}}<<E_{\rm{Bp}}$. From Eq.
(\ref{tEq13}), $\partial_xP_l^s$ has the order of magnitude $\sqrt{\zeta q |{\rm{Re}}(A_1)|}$ in the narrow region $1/\sqrt{\zeta q |{\rm{Re}}(A_1)|}$. Thus, the order of
$E_{\rm{Bs}}$ is $\sqrt{|q{\rm{Re}}{(A_1)}|/\zeta}<<1$. On the other hand, $E_{\rm{Bp}}$ is proportional to $(A_1k_1)^2$, which is of order of $10$ with the parameters used in the
present system. This also indicates that the magnetic energy for different kinks is roughly the same (thus we only show the magnetic energy for one kink configuration in Fig.
\ref{f7} (a)). It is quite different from the usual solitons in a single Josephson junction.

With the linear expansion of Josephson current, similar to Eq. (\ref{tEq9}), the Josephson energy $E_{\rm{J}}$ in the state with kink can be obtained
\begin{equation}\label{sEq2}
    E_{\rm{J}}=1-\frac{(k_1^2-\omega^2)|F_1|^2/4}{(k_1^2-\omega^2)^2+\beta^2\omega^2}.
\end{equation}
It first decreases and then increases inside the current step, while is close to unity off resonance. The calculation of the electric energy and total energy is straightforward.
The results are shown in Figs. \ref{f7} (b) and (d). The total energy for different kinks is approximately the same. Therefore the states with kink occupy finite volumes in the
phase space with the same energy, which makes this state easily accessible.

In the state without kink, the magnetic energy is obviously zero. The Josephson coupling is
\begin{equation}\label{sEq3}
    E_{\rm{J}}=1+\frac{1}{2(\omega^2+\beta^2)}.
\end{equation}
It decreases from its maximum at the retrapping point and saturates to unity at large currents, which is consistent with the results shown in Fig. \ref{f7}(c). $E_{\rm{J}}$ in the
region of $J_{\rm{ext}}<J_{\rm{r}}$ is very close to $0$, which cannot be described by Eq. (\ref{sEq3}) because the system is retrapped into superconducting state. The total
energy is the same as that of state with kink in the linear ohmic \emph{IV} curve, while it is larger than that of state with kink in the current step.

The electric energies in Fig. \ref{f7} is much larger than other energy in the state without kink in a sharp contrast to cases at equilibrium, where different energy contributions
are expected to be the same. The reason can be understood if we consider Eq. (\ref{eq5b}) as the equation of motion of phase particle in a titled washboard potential. In the
presence of external current, the phase particle is accelerated and start to run in the tilted washboard potential. In response to the modulated potential, small oscillations of
phase particle are created in addition to the motion with a constant velocity. Meanwhile, the motion of phase particle causes dissipation. The steady state is reached when the
input power and dissipation are balanced. On the other hand, the magnetic energy and Josephson coupling are solely contributed from the small oscillation of the phase particle. As
the most part of input power converted into the motion with a constant velocity, the electric energy occupies the most part of energy stored in the system, as shown in Fig.
\ref{f7}. However, in the state with kink, as a significant portion of the input power is converted into plasma oscillation in the current steps, rather than to solely increase
the velocity of the  phase particle, the electric energy and magnetic energy become comparable to each other.
\begin{figure*}[t]
\psfig{figure=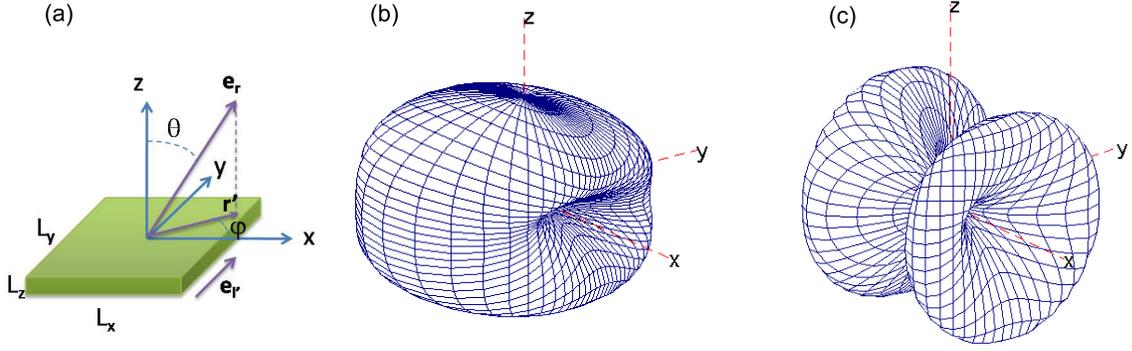,width=18cm} \caption{\label{f7b}(Color online). (a) Coordinate system for the calculation of
radiation pattern from the mesa. (b) Radiation pattern of the mode $(1,0)$. (c) Radiation pattern of the mode $(1,1)$.
Here $L_x=80\rm{\mu m}$, $L_y=300\rm{\mu m}$ and $L_z=1\rm{\mu m}$.}
\end{figure*}

\begin{figure}[b]
\psfig{figure=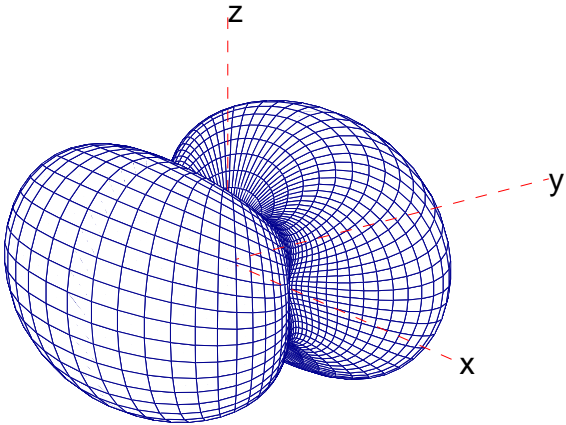,width=7.5cm} \caption{\label{f7f}(Color
online). Radiation pattern of the state without kink biased at the
retrapping point. Here $L_x=80\rm{\mu m}$, $L_y=300\rm{\mu m}$ and
$L_z=1\rm{\mu m}$. The anisotropy of the pattern in the $xy$ plane
is due to the fact $L_y>>L_x$.}
\end{figure}

\section{Radiation pattern}
In this section, we calculate the far-field radiation pattern for the mesa operated in the state with kink and without kink, which is important both for applications and for
differentiating various states.

To calculate the radiation pattern, we resort to the Huygens principle in which the pattern is determined by the
oscillation of the electromagnetic fields at the edges of samples, which can be casted into the edge magnetic current
and electric current in the formula of equivalence principle.\cite{StutzmanBook} Since there exists a significant
impedance mismatch, the electric current produced by the oscillating magnetic field is much smaller than the magnetic
current produced by the oscillating electric field, so we can neglect the contribution from the electric current. The
equivalent magnetic current $\mathbf{M_e}$ in the dimensionless units is given by
\begin{equation}\label{eqpat1}
\mathbf{M_e}=\mathbf{E_e}\times\mathbf{n},
\end{equation}
where $\mathbf{E_e}$ is the oscillating electric field at the edges of mesa and the vector $\mathbf{n}$ is normal to
the edges. As we know that the radiation pattern critically depends on the geometry of the source, we need to consider
the $3$D system. The extension from previous analysis of $2$D system to $3$D is given in Ref.\cite{Hu08}. The
coordinates for the $3$D system are sketched in Fig. \ref{f7b}(a). We use the similar dimension as in the
experiments,\cite{Ozyuzer07, kadowaki08} i.e. $L_x=80\rm{\mu m}$, $L_y=300\rm{\mu m}$ and $L_z=1\rm{\mu m}$. Because
$k_{\omega}L_z<<1$ with $k_{\omega}\equiv \omega/c$, the sources at different $z$ coordinates do not interfere too
much, and can be treated as uniform. In this case, the far-field Poynting vector in the dimensionless units is
\begin{equation}\label{eqpat2}
\mathbf{S_r}=\frac{\omega^2L_z^2}{32\pi^2r^2\varepsilon^{3/2}}|\mathbf{G}|^2\mathbf{e_r},
\end{equation}
with
\begin{equation}\label{eqpat2b}
\mathbf{G}=\oint_{\rm{edges}}M_e(r')\exp(-i\frac{\omega}{\sqrt{\varepsilon}}\mathbf{r'}\cdot\mathbf{e_r})(\mathbf{e_r}\times\mathbf{e_{l'}})d\mathbf{{l'}},
\end{equation}
where the integral is taken over the perimeter of the crystal.\cite{Leone03,Koshelev08} With the size we use, the
interference is mainly contributed from the source along the y direction since $L_y$ is comparable to the wavelength.

\begin{figure}[t]
\psfig{figure=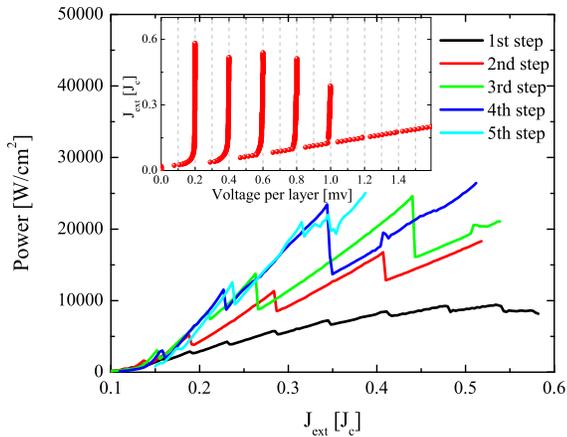,width=\columnwidth} \caption{\label{f8}(Color online). Radiation power from the zero-field steps caused by soliton motions. The inset is the \emph{IV}
characteristics. The vertical dashed lines are the assignment of cavity mode according to $k_n=n\pi/L$ with $n$ being an integer and the ac Josephson relation. The results are
obtained with $C=0.000177$ and $R=707.1$.}
\end{figure}

\begin{figure}[b]
\psfig{figure=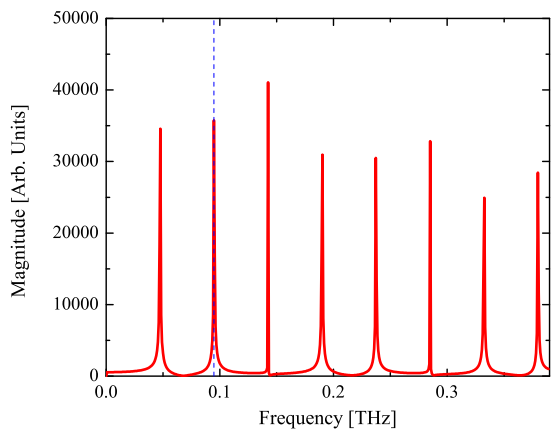,width=\columnwidth} \caption{\label{f9}(Color online). Frequency spectrum at the first zero-field step at $J_{\rm{ext}}=0.57$. The vertical dashed line is
the frequency given by the ac Josephson relation with voltage $V=2\pi/L$. The fundamental peak does not come to the dashed line means that the ac Josephson relation is broken. The
results are obtained with $C=0.000177$ and $R=707.1$.}
\end{figure}

In the state with kink, the oscillation of the electric field in the frequency domain can be well described by
\begin{equation}\label{eqpat3}
E_z=A_1\omega\cos(n_x\pi/L_x)\cos(n_y\pi/L_y),
\end{equation}
for the cavity mode $(n_x, n_y)$ when the plasma oscillation is weak.\cite{Hu08} The radiation pattern from the mode $(n_x, n_y)$ can be evaluated with Eq. (\ref{eqpat2}) and is
reported in Refs. \cite{Leone03,Koshelev08}. Inside the current step, the higher harmonics become important so numerical simulations are needed. We use computer simulations to
calculate the oscillation of electric field at edges and then substitute the results into Eqs. (\ref{eqpat1}) and (\ref{eqpat2}) to obtain the radiation pattern. The results at
current steps corresponding to the cavity modes $(1,0)$ and $(1,1)$ are shown in Figs. \ref{f7b} (b) and (c). For mode $(1,0)$, the radiation power is maximal at the top of the
mesa $\theta=0$, and it has a maximum at the middle of $L_x$ while a minimum at the middle of $L_y$; for mode $(1,1)$, the radiation power is minimal at the top of the mesa
$\theta=0$, and at the middle of $L_x$ and $L_y$.

In the state without kink, the oscillation of the electric field is homogenous in the $xy$ plane, which corresponds to the $(0, 0)$ mode. The radiation pattern at the retrapping
point is presented in Fig. \ref{f7f}. It has a minimum both at the top of the mesa and at the middle of $L_y$, and a maximum at the middle of $L_x$. The anisotropy of the pattern
in the $xy$ plane is due to the fact $L_y>>L_x$.

\section{state with solitons}
To be comprehensive, we present here the results of numerical simulations on the state with solitons. It should be noted that in the present system, the length scale is
$\lambda_c$ rather than $\lambda_J$ as in conventional Josephson junctions and in the presence of magnetic field. Therefore, to have solitons, the length of the junctions must be
larger than $\lambda_c$. Because of repulsive interaction, it is believed to be hard to achieve in-phase motion of solitons in a stack of junctions, despite some simulations
suggest that the solitons in high velocity have attractive interaction.\cite{Krasnov97,Gorria03} Here we investigate the radiation due to soliton motions in a single junction,
which is equivalent to a stack of junctions if one realizes the in-phase motion of solitons in different junctions.

It is well known that periodic motions and reflections of solitons and anti-solitons give birth to the zero-field steps at $V=2n\pi/L$, which corresponds to the even cavity
modes,\cite{Fulton73,Ustinov98} with $n$ the number of solitons. When a soliton hits the boundary, it emits an electromagnetic \emph{pulse}.\cite{Lomdahl82} Here we only
investigate the radiation from zero-field steps and will not discuss the Cherenkov radiation.\cite{Hechtfischer97} We perform computer simulations to trace out all the zero-field
steps. We use $L=5\lambda_c$ so that there exist five steps. The \emph{IV} characteristics is shown in the inset of Fig. \ref{f8}, where the zero-field steps occur at the voltage
corresponding to the even cavity modes. The radiation power at each step is shown in Fig. \ref{f8}, which is higher than that from the state with kink if one assumes in-phase
motion of solitons. The discontinuous drops in the radiation power when ramping up the current are caused by the change in the wavelength of the Josephson plasma excited by the
motion of solitons. The frequency spectrum at the first zero-field step is sketched in Fig. \ref{f9} and there are many frequency harmonics with the fundamental frequency not
satisfying the ac Josephson relation. In the state with solitons, the radiated frequency depends on the configuration of solitons except the first zero-field step.\cite{Dueholm81}
It is noticed that, as indicated in Fig. \ref{f9}, the fundamental frequency and voltage at all the steps never satisfy the ac Josephson relation, contrasting with the state of
kink revealed theoretically\cite{szlin08b} and the experimental observations\cite{Ozyuzer07,kadowaki08}.

\section{Discussions and Conclusions}
In the present work, the phase dynamics and its electrodynamics in a thick stack of intrinsic Josephson junctions in the absence of external magnetic field are investigated both
analytically and numerically. There is good consistency between the analytical theory and simulations.

In the state with phase kink, there are many current steps at both even and odd cavity modes. The phase kink plays a role of coupling the plasma to the cavity modes, as such the
plasma oscillation is largely enhanced. The radiation power from the state with phase kink is $\omega^4$ times larger than that without phase kink. The plasma oscillation is
uniform through the $c$ axis. Thus the far-field radiation power grows as $N$ squared, the so-called superradiation. At the bottom of the first current step, the magnetic field is
symmetric with respect to the center of the junction. The antisymmetric component becomes more and more important when going into the current step. The states with different phase
kink configurations are degenerate in the sense that they have the same \emph{IV} characteristics, power radiation and energy stored in the system.

In the state without phase kink, the plasma oscillation does not couple to the cavity mode, and thus the radiation power is very small. The power increases with decreasing current
and reaches the maximum at the retrapping point. The radiation occurs in a broad region of voltage. The frequency satisfies the ac Josephson relation, and high frequency harmonics
are almost invisible. The magnetic field is antisymmetric in this state. The far-field radiation pattern in this state is quite different from that in the state with phase kink,
which is a clear fingerprint of the dynamic state realized by the system.

In the state with solitons, electromagnetic pulses are radiated from junctions when solitons hit the boundary. There are many frequency harmonics, but the fundamental frequency
never satisfies the ac Josephson relation. It would be ideal for exciting strong terahertz wave with solitons because the power is about $25000\rm{W/cm^2}$, presuming one could
realize the in-phase motion of solitons in a thick stack of long intrinsic Josephson junctions.

It is illuminating to discuss the dynamic state realized in the recent experiments for terahertz radiation\cite{Ozyuzer07,kadowaki08} in light of the present theoretical analysis.
Coherent radiations were detected in the resistive curve in Ref. \cite{Ozyuzer07} with the frequency corresponding to the first cavity mode. One order of the magnitude stronger
radiations were observed in the region of voltage with anomalous \emph{IV} characteristics in Ref. \cite{kadowaki08}, and there are many frequency harmonics at a given voltage. In
both experiments, the frequency obeys the ac Josephson effect and thus the state with traveling solitons can be ruled out. On the other hand, large cavity resonances cannot be
excited in the state without phase kink; furthermore, the radiation from the state without phase kink occurs weakly in a wide range of voltage. Therefore, it is unlikely relevant
to the experimental observations. The state with phase kink, in contrast, seems to be consistent with the experiments so far. In this state, the plasma oscillation is uniform
through the stack of Josephson junctions, it thus supports superradiation as observed in the experiments. Moreover, the periodic arrangement of static kink along the stack
direction allows to pump dc powers into large plasma oscillations, which yields self-induced current steps. It is noticed that, to obtain the overall shape of the \emph{IV} curve,
we need to take the heating effect into account. More works are needed to clarify the synchronization process.

It should be remarked that there are propagating waves besides the standing wave in Eq. (\ref{tEq1}), because of the radiation. The amplitude of the propagating waves has the
order of magnitude of $1/|Z|$ and thus can be safely neglected as the first-step approximation, which is confirmed by the numerical simulations. The radiation has only some
negligible effects on the dynamics inside the junctions, which permits us to calculate the radiation perturbatively. When the mismatch of impedance at the edges is reduced, e.g.,
the thickness of a stack of intrinsic Josephson junctions is comparable to $\lambda_c$, one has to consider the radiation and interference for the analysis of phase dynamics
inside the junctions self-consistently.

The essential property of the stack of junctions for realization of the state with kink is the strong inductive
coupling. As for other possible effects on the state with phase kink, we find in the simulations that this state is
very stable against small magnetic fields, thermal fluctuations.\cite{szlin08c} In Ref.~\cite{Koshelev08b}, it is shown
that the state with kink is stable against the modulation of critical current. Thus it is likely to be realized
experimentally. The state with kink is promising for the application of terahertz radiation. It is also useful for
terahertz detectors, amplifiers and mixers.

\section{Acknowledgement}
The authors thank U. Welp, K. Kadowaki, M. Tachiki, L. Bulaevskii, A. Koshelev, N. Pederson, H. B. Wang and T. Koyama for helpful discussions. They are also indebt to K. Kodowaki
for showing them the experimental results prior to publication. Calculations were performed on SR11000 (HITACHI) in NIMS. This work was supported by WPI Initiative on Materials
Nanoarchitronics, MEXT, Japan, CREST-JST, Japan and partially by ITSNEM of CAS.


\end{document}